\documentclass[twocolumn,showpacs,preprintnumbers,amsmath,amssymb,footinbib]{revtex4}

\usepackage{graphicx}
\usepackage{dcolumn}
\usepackage{bm}

\begin{document}

\preprint{APS/123-QED}
\title{Nernst Effect in NdBa$_2$\{Cu$_{1-y}$Ni$_y$\}$_3$O$_{7-\delta}$}

\author{N. Johannsen$^{1*}$, Th. Wolf$^2$, A. V. Sologubenko$^1$, T. Lorenz$^1$, A. Freimuth$^1$ and J. A. Mydosh$^1$}

 \affiliation{$^1$II. Physikalisches Institut, Universit\"{a}t zu K\"{o}ln, Z\"{u}lpicher Str. 77, 50937 K\"{o}ln, Germany\\
$^2$Forschungszentrum Karlruhe, Institut f\"{u}r Festk\"{o}rperphysik, 76021 Karlsruhe, Germany
}

\date{\today}

\begin{abstract}
In NdBa$_2$\{Cu$_{1-y}$Ni$_y$\}$_3$O$_{7-\delta}$, magnetic Ni-impurities suppress $T_{c}$ but at the same time the pseudogap is strongly enhanced. This unique feature makes it an ideal system to study possible relations between the anomalous Nernst effect, superconductivity and the pseudogap. We present Nernst effect measurements on a series of optimally doped (O$_{7}$) and underdoped (O$_{6.8}$) samples with Ni contents ranging from y=0 to 0.12. In all samples an onset of the Nernst signal is found at $T^\nu$$>$$T_c$. For the optimally doped samples $T^\nu$ and $T_c$ decrease simultaneously with increasing Ni content. The underdoped samples show a different behavior, i.e. the onset of the Nernst signal is hardly affected by increasing the Ni content from y=0 to 0.03. Irrespective of the oxygen content, $T^\nu$ clearly does not track the enhanced pseudogap temperature $T^*$.
\end{abstract}

\pacs{74.40.+k, 72.15.Jf, 74.62.Dh, 74.72.-h}

\maketitle

More than 20 years after the discovery of high-temperature superconductivity the mechanism is still unsolved. Possible relations to other anomalous phenomena in high-$T_c$ materials such as the pseudogap phase may play a key role towards an understanding of this mechanism.
The pseudogap is exhibited in the normal-state region of the underdoped compounds. First evidence for a pseudogap in the charge channel was reported on the basis of optical conductivity data \cite{homes93}, extending to temperatures $T^*$\,$\gg$\,$T_c$. It has been suggested that this phenomenon is related to superconductivity via preformed Cooper pairs that are thermally prevented from forming long-range phase coherence \cite{emery95}. Other suggestions include novel phases that may even compete with superconductivity, such as exotic spin or charge density wave states \cite{chubukov98, affleck88, wen96, chakravarty01} and stripe ordering scenarios \cite{low94}.
The preformed-pair scenario seems to be the most attractive description because it corresponds to the three-dimensional equivalent of the well-known two-dimensional Kosterlitz-Thouless (KT) transition in which thermally excited vortices (in zero field) destroy phase rigidity but not the pairing amplitude.

Anomalous Nernst signals that are detectable far into the pseudogap region as well as a diamagnetic response which scales with the Nernst effect support this description \cite{wang06}. Since a large Nernst effect in the mixed state of type-II superconductors is attributed to moving vortices \cite{xu00} it is a very accurate probe for vortex formation and vortex-like excitations. The Nernst signal is anomalous if it extends into the normal-state region because a charge carrier Nernst signal is usually very small in normal metals due to the so called ``Sondheimer cancellation'' \cite{sondheimer48}. In clean high-$T_c$`s with moderate inelastic scattering the carrier-Nernst signal can reach values of about 25\% of the magnitude of the vortex-Nernst signal but it is linear in field and opposite in sign to the vortex-signal \cite{oganesyan04}.
Thus it is obvious that Nernst effect measurements can effectively contribute to unravel and clarify the pseudogap phenomena. 

In order to facilitate such investigations one should examine a system in which the most relevant parameters, $T_c$ and $T^*$, can be tuned individually. 
Given such a system, one could examine the properties of the anomalously enhanced Nernst signals that are found above $T_c$ and below $T^*$. In order to determine if there is a connection to the pseudogap, one just has to shift the pseudogap and observe whether the Nernst signal follows or not. 
Fortunately, such a system exists, namely NdBa$_2$\{Cu$_{1-y}$Ni$_y$\}$_3$O$_{7-\delta}$. Recent measurements of the optical conductivity of Ni-doped NdBa$_2$Cu$_3$O$_{7-\delta}$ revealed a strongly enhanced pseudogap while superconductivity is suppressed at the same time \cite{pimenov05}. This compound therefore offers the unique opportunity to study the development of the Nernst signal while tuning $T_c$ and the pseudogap individually and even in opposite directions.
Our measurements were performed on a series of optimally doped (O$_7$) and underdoped (O$_{6.8}$) samples with Ni contents ranging from y = 0 to 0.12. Irrespective of the oxygen content, the onset of the Nernst signal ($T^\nu$) clearly does not follow the enhanced pseudogap.

\begin{figure*}[ht]
\includegraphics[width=16cm, height=5cm]{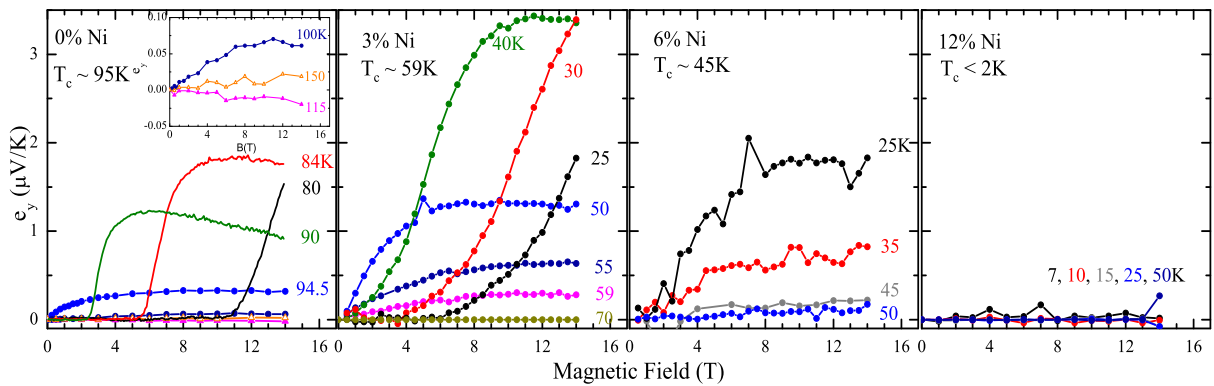}
\caption{\label{fig:nernst_vergleich_optimal_mag_inset} (color online) Typical Nernst signals of a series of optimally doped NdBa$_2$\{Cu$_{1-y}$Ni$_y$\}$_3$O$_{7}$. The inset enlarges the contributions to the Nernst signal above $T_c$.}
\end{figure*}

\begin{figure}[hb]
\includegraphics[width=8cm, height=5cm]{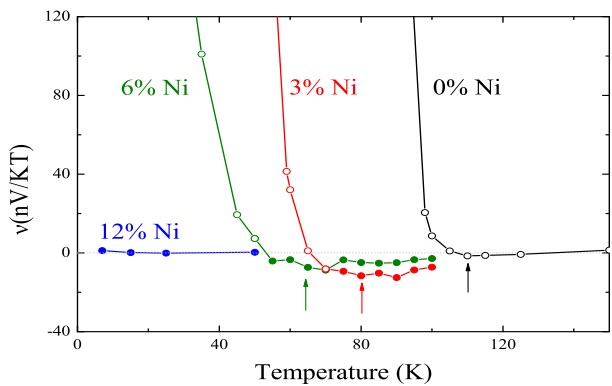}
\caption{\label{fig:onset_optimally_doped} (color online)  Nernst coefficients $\nu$ versus temperature for the pure and three Ni-doped (O$_7$) single crystals \cite{footnote}. Onset temperatures are indicated by arrows.}
\end{figure}

The high-quality single crystals of NdBa$_2$\{Cu$_{1-y}$Ni$_y$\}$_3$O$_{7-\delta}$ used in our study were grown by a flux method as described in detail in \cite{pimenov05}. Ni substitution does not alter the hole doping content significantly which is confirmed by thermopower measurements. It is estimated that half of the Ni impurities reside within the CuO$_2$ planes. The influence of substituting magnetic Ni ions or nonmagnetic Zn ions in the CuO$_2$ planes has been investigated by several groups \cite{kuo97,hudson01,xu05}. In YBa$_2$Cu$_3$O$_{7-\delta}$ (YBCO) a linear depression of $T_c$ is found for both impurities \cite{kuo97}.
STM measurements revealed the local electronic structure for both substitutes \cite{hudson01}. Ni retains a local magnetic moment and the superconducting energy gap is unimpaired. Scattering at this site is then mostly determined by potential interactions. 
The effects of impurity doping on the onset of the Nernst effect has been studied by several groups. In Zn-doped YBa$_2$Cu$_3$O$_{7}$ $T^\nu$ was found to decrease with $T_c$ while the pseudogap is left untouched \cite{xu05}. In electron irradiated samples of YBa$_2$Cu$_3$O$_{7}$ and YBa$_2$Cu$_3$O$_{6.6}$ a constant onset of the Nernst effect was reported for both samples \cite{rullier-albenque06}, but the authors did not address the relation between the Nernst effect and the pseudogap.
Depending on the Ni-impurity level in NdBa$_2$\{Cu$_{1-y}$Ni$_y$\}$_3$O$_{7-\delta}$, we can compare the enhanced pseudogap to $T_c$ and $T^\nu$.

The Nernst effect measurements were carried out on four optimally doped NdBa$_2$\{Cu$_{1-y}$Ni$_y$\}$_3$O$_{7}$ samples. The $T_c$ values and transition width, $\Delta$$T_c$ are taken as the midpoint of the diamagnetic transition and the 10-90\% width of the transition, respectively. A pure sample with $T_c$\,=\,95($\Delta T_c$\,=\,1)\,K, and three samples with y\,=\,0.03~[$T_c$\,=\,59(8)\,K], 0.06~[$T_c$\,=\,45(10)\,K] and 0.12 with $T_c$$<$\,2\,K. 
By annealing in flowing O$_2$ atmosphere at about 540$^{\circ}$C and P=1\,bar a series of underdoped (O$_{6.8}$) samples were prepared: y\,=\,0~[$T_c$\,=\,53(7)\,K], 0.03~[$T_c$\,=\,20(14)\,K] and 0.06 with $T_c$$<$\,2\,K. Typical sample sizes were 1\,mm$\times$1\,mm$\times$0.5\,mm. For the Nernst effect measurements two thin copper wires were attached at the sides of the sample with a two-component silver epoxy in order to pick up a transverse voltage. The sample was then mounted to a copper block which was at the same temperature as the temperature-stabilized surrounding. On the top of the sample a chip heater was thermally connected by an insulating varnish (VGE 7031, Lakeshore). By applying power to the heater a temperature gradient is created over the sample of about 0.5\,K/mm which is detected by a AuFe-Chromel thermocouple. Two methods were used to detect the Nernst signals. Most of the measurements were performed by stabilizing the temperature (to about 1\,mK) and the temperature gradient for each magnetic field (typically in steps of 0.5\,T). Then the temperature gradient is removed in order to subtract offset voltages. For the second method  the temperature is stabilized again very accurately (1\,mK), then the heater power is switched on and the field is continuously swept at a typical rate of about 0.3\,T/min. A second sweep with the heater switched off is performed at the same temperature in order to remove field-dependent offset voltages. Both methods give excellent agreement. The Nernst signal itself is extracted as the antisymmetric part of the field dependent transverse voltages in order to eliminate thermopower voltages due to possible misalignments of the contacts and is given by
\begin{equation}
e_y=\nu B_z=\frac{E_y}{-\nabla T_x},
\end{equation} 
where $\nu$ is the Nernst coefficient and $E$ denotes the electric field.
\begin{figure*}[ht]
\includegraphics[width=16cm, height=5.5cm]{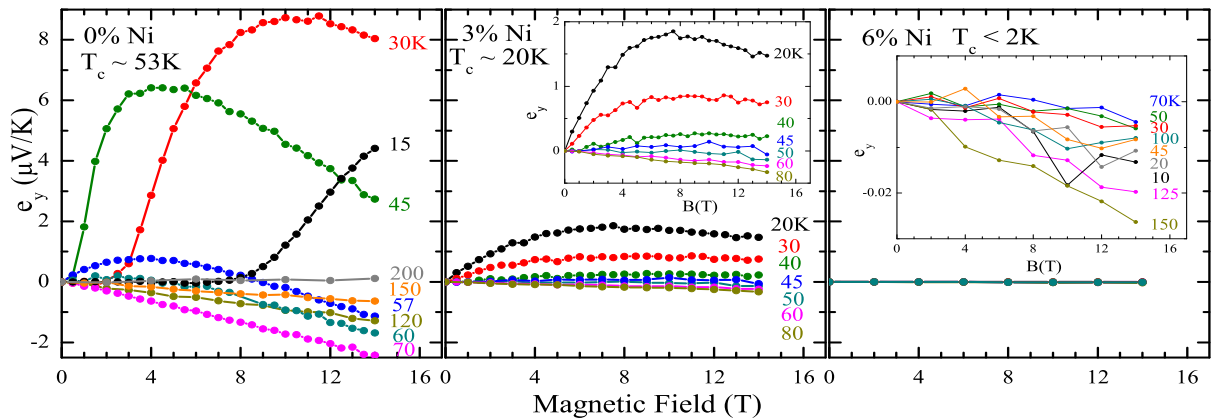}
\caption{\label{fig:nernst_vergleich_underdoped_mag} (color online) Nernst signal of a series of underdoped NdBa$_2$\{Cu$_{1-y}$Ni$_y$\}$_3$O$_{6.8}$. The insets show an expanded view of the data.}
\end{figure*}

\begin{figure}[hb]
\includegraphics[width=8cm, height=5cm]{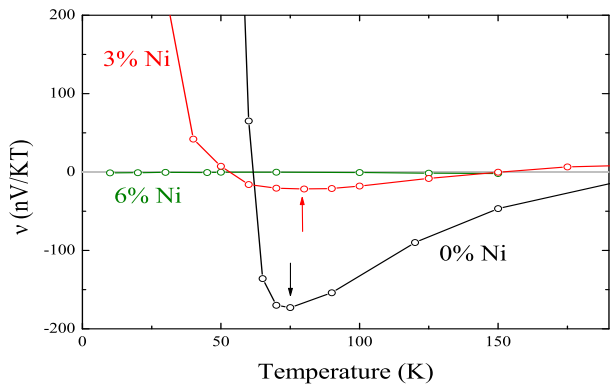}
\caption{\label{fig:v_onset_0_3_underdoped}(color online)  Nernst coefficients $\nu$ of the underdoped compounds (Onset temperatures are marked by arrows).}
\end{figure}

 Figure \ref{fig:nernst_vergleich_optimal_mag_inset} shows measurements of the optimally doped samples for Ni contents of 0, 3, 6 and 12.\%. At low T, the signal is zero until a typical temperature-dependent melting field is reached. At that field the density of vortices is sufficiently high to overcome pinning forces. In the adjacent ``vortex-liquid'' phase the vortices are free to move towards the cold end of the sample and thereby are able to produce the phase-slip voltage that gives the anomalously enhanced Nernst signal. This causes the steep increase in the Nernst signal. When the field is further increased, the signal saturates until it slowly diminishes towards the upper critical field H$_{c2}$. This results in the characteristic ``tilted-hill'' profile of the Nernst signal in high-$T_c$ cuprate superconductors \cite{wang03}. Now increasing the temperature and approaching $T_c$ the melting field begins to weaken until the signal rises immediately at H=0. Above $T_c$ the signal drops very sharply until it becomes very small, negative and linear in field.

The linear and negative contribution in the normal state that can be identified with the quasi-particle (QP) background \cite{wang01,oganesyan04} which is very small in this optimally doped compound. This small background seems to be a common feature of the optimally doped [Y,Nd]-BCO  samples \cite{rullier-albenque06, wang06, xu05}, independent of the impurity-content. 
Increasing the Ni content to 3\%, the absolute values of the signals increase while $T_c$ drops to 59(8)\,K. In contrast to the sharp increase of the signal after exceeding the melting field in the pure sample, the signal rises more smoothly in the 3\% Ni-doped compound. By further increasing the Ni-content to 6\% the absolute values of the Nernst signal decrease again until the signal can not be distinguished from the background at 12\% Ni. In the entire series of the optimally doped samples the Nernst signal above $T_c$ drops quickly. As mentioned above the QP contribution is very small.

Figure \ref{fig:onset_optimally_doped} shows the Nernst coefficient, $\nu=e_y/B_z$. Dividing by the field yields a field independent value only for the linear contributions of the Nernst signals produced by the QP background. At lower T, a voltage due to vortex movement adds to the QP contribution which leads to a strong field dependence of $\nu=\nu(B)$. In Figure \ref{fig:onset_optimally_doped} the open symbols are derived by linearly fitting the initial slope of $e_y$ \cite{footnote}. The onset of the Nernst signal is then defined as the most negative value of $\nu$.
In Ref.~\cite{pimenov05} it is reported that the pseudogap can be restored or even enhanced by adding Ni-impurities to the optimally doped (or slightly overdoped) samples. This enhanced pseudogap can now be compared to the onset of the Nernst signals. 
The onset, $T^\nu$ of Figure \ref{fig:onset_optimally_doped} (indicated by arrows) shifts to lower temperatures as the Ni content is increased. If there was a scaling of $T^\nu$ with the pseudogap temperature one would expect the opposite behavior since the sample with 12\% Ni is identified with the highest pseudogap temperature \cite{pimenov05}. Here the decrease of  $T^\nu$ is, however, closely connected to the decrease of $T_c$. 

\begin{figure}
\includegraphics[width=8cm]{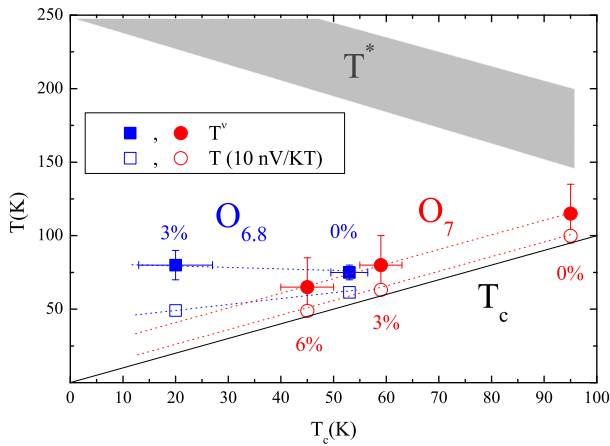}
\caption{\label{fig:phase_diagram_onset_vs_tc} (color online) Onset temperatures of all samples versus $T_c$. The $T^*$ band sketches the pseudogap opening temperatures as estimated from the results of \cite{pimenov05}.}
\end{figure}

Next, we consider the behavior of the underdoped compounds at Ni-impurity levels of 0, 3 and 6\% (Figure \ref{fig:nernst_vergleich_underdoped_mag}). The pure sample shows the largest vortex contribution to the Nernst signal of all measured samples. In addition, a very large negative QP signal is present which exceeds -2\,$\mu$V/K at 70\,K and 14\,T. 
Adding Ni, the anomalous Nernst signal as well as the QP contribution decrease very sharply (3\%) until the signal is further diminished and no systematic temperature dependence of the negative background can be deduced (6\%). 
Surprisingly, the onset temperatures of both underdoped samples do not change with the Ni content (Figure \ref{fig:v_onset_0_3_underdoped}). This result differs from the onsets of the optimally doped series. 
 All onset temperatures of our samples are collected in Figure \ref{fig:phase_diagram_onset_vs_tc} as a function of $T_c$. These onsets can now be compared to the pseudogap data of \cite{pimenov05}. 
 The fluctuation regime is found to be narrow in our pure samples, e.g., the onsets are determined to be about 20\,K higher than $T_c$. The scaling behavior of $T^\nu$ changes from the O$_7$ to the O$_{6.8}$ compounds. Although $T^\nu$ of the optimally doped samples clearly follow $T_c$, the onsets of the  underdoped ones seem to be independent of the Ni content. But clearly no scaling to the reported enhanced pseudogap can be established for both series since it is shown that the pseudogap and the pseudogap temperature increase with increasing Ni-content \cite{pimenov05}.
Another criterion is the temperature which matches a slightly positive Nernst coefficient of $\nu=10$\,nV/KT \cite{wang06,rullier-albenque06}. T($\nu=10$\,nV/KT) (open circles) scales downward from $T^\nu$ and is thereby proportional to $T_c$ in the optimally doped compounds. The underdoped compounds develop a slope T($\nu=10$\,nV/KT) that tilts towards the slope of $T_c$ (open squares) but is still clearly distinguishable from the optimally doped ones.

A sharp decay of the Nernst signal above $T_c$ is predicted by Ref.~\cite{podolsky06} whose calculations yield a proportionality $T^\nu$$\sim$\,$T_c$. This result is in agreement with our findings from the optimally doped samples.  The high onset with respect to $T_c$ in the underdoped phase (3\%\,Ni) is comparable to the high onsets found in electron-irradiated YBCO$_{6.6}$ samples \cite{rullier-albenque06} as well as in Y$_{1-x}$Pr$_x$Ba$_2$Cu$_3$O$_{7-\delta}$ \cite{li07}. These can be interpreted in terms of an unchanged $T^\nu$ while  $T_c$ is suppressed due to the impurity-induced disorder. 
Recent calculations connect the onset of the Nernst signal to the onset of superconductivity in spatially disordered systems \cite{dias06}. This onset is then found to be at temperatures that represent the onset of superconducting islands, which percolate at $T_c$. This implies $T^\nu$$\ll$\,$T^*$, which coincides with our findings.

 In summary, we have studied the Nernst effect for a series of optimally doped and underdoped NdBa$_2$\{Cu$_{1-y}$Ni$_y$\}$_3$O$_{7-\delta}$ samples with Ni concentrations varying between y=0 and 0.12. Substituting Ni for Cu gives the unique opportunity to study the development of the Nernst effect in a system where the pseudogap temperature $T^*$ and the superconducting $T_c$ can be tuned separately.
For the optimally doped samples a clear scaling ($T^{\nu}$$\sim$20\,K+$T_c$) exists while in the underdoped ones $T^{\nu}$ remains Ni-independent. For both series, there is no correlation of $T^\nu$ with the enhanced pseudogap.
\begin{acknowledgements}
We acknowledge fruitful discussions with M. Vojta and D. I. Khomskii.
This work was supported by the Deutsche Forschungsgemeinschaft through SFB~608.
\end{acknowledgements}

*Electronic address: johannse@ph2.uni-koeln.de
\end{document}